# Spontaneous symmetry breaking in synchronously pumped fiber ring cavities

M. J. Schmidberger[1,2,*], D. Novoa[1], F. Biancalana[1,3], P. St.J. Russell[1,2] and N. Y. Joly[2,1]

[1]*Max Planck Institute for the Science of Light, Günther-Scharowsky-Str. 1, 91058 Erlangen, Germany*
[2]*Department of Physics, University of Erlangen-Nuremberg, Staudtstraße 7, 91058 Erlangen, Germany*
[3]*School of Engineering & Physical Sciences, Heriot-Watt University, Edinburgh EH14 4AS, UK*
*[*]michael.schmidberger@mpl.mpg.de*

We introduce a new equation that describes the spatio-temporal evolution of arbitrary pulses propagating in a fiber-ring cavity. This model is a significant extension of the traditionally used Lugiato-Lefever model. We demonstrate spontaneous symmetry breaking as well as multistability regimes in a synchronously pumped fiber-ring cavity. The equation can be applied to virtually any type of waveguide-based ring cavity.

PACS numbers: 05.45.-a, 11.30.Qc, 42.65.-k

Spontaneous symmetry breaking (SSB) mechanisms are known to be ubiquitous in nature. In particular, they are responsible for a plethora of paradigmatic physical phenomena, ranging from ferromagnetism and superconductivity [1] to fundamental predictions like the celebrated Higgs mechanism [2]. In the fields of nonlinear photonics and matter-wave dynamics, SSB is usually attributed to the combined effect of linear waveguiding structures and nonlinear interactions (see e.g. [2–6] and references therein), and generates an asymmetry in the profiles of stationary states of the system.

Both linear and nonlinear features are present in waveguide-ring cavities, making such systems ideal for studying SSB. Passive fiber ring cavities are simple devices that can exhibit extremely rich nonlinear dynamics, as first reported by Kensuke [7] and thoroughly investigated over the past three decades [8–12]. From the experimental point of view, the advent of photonic crystal fiber (PCF) as a nonlinear element has sparked new interest in this established field due to the possibility of adjusting dispersion and nonlinearity into previously inaccessible parameter ranges [13–15]. Two approaches have previously been used for theoretical modeling of such cavities: the Ikeda-map formalism [7] (IMF) and the Lugiato-Lefever (LL) approach [16]. In the IMF, the evolution of the pulse during each trip around the cavity is modeled numerically in discrete steps. In this frame, all physical phenomena such as dispersion, nonlinearity and losses are applied sequentially to the field.

Commonly used with pulsed sources, this approach allows for arbitrary pump pulse durations and shapes as well as large values of temporal walk-off $\tau$ between the round-trip time and the pump pulse repetition period [14]. In contrast, the LL model relies on a single partial differential equation for the mean field, and is hence computationally much less demanding and more amenable to analysis. Although it is, in principle, restricted to continuous pumping and $\tau$ values of order of an optical cycle (which we define as interferometric), it has also been successfully applied in quasi-continuous cases [17]. Recently, partly as a result of some important applications in the field of supercontinuum generation [18], there has been a resurgence of interest in fiber-based cavities, making it necessary to extend the LL equation to a more realistic model capable of handling ultrashort pulses [19]. Third order as well as higher order dispersion terms have been included in the LL equation, in order to explain, e.g., the observation of bistability and the emergence of temporally asymmetric pulse shapes in a fiber ring cavity [20–22]. A similar extension of the LL equation has been proposed to study other schemes such as whispering gallery mode resonators [23].

In this Letter we propose, following a multiscale approach, a novel and very general LL-like equation that is capable of dealing with ultrashort pulses and can support walk-off values of order the pulse duration. As examples, we apply this equation first to model the dynamics of an asynchronously pumped solid-core PCF ring cavity, and then to study the multistable operation of a gas-filled hollow-core PCF ring cavity. In the second case the model reduces to a fully temporally inversion-symmetric equation, which remarkably predicts the spontaneous emergence of asymmetric pulse shapes, thus exhibiting a kind of SSB. Moreover, for certain parameter ranges asymmetric solutions are the only stable solutions of the system. This can be explained using a hydrodynamical representation of the internal energy flow of the pulse within the cavity.

This proposal relies on a multiscale analysis similar to the well known Haus master equation [24], successfully used to study the dynamics of mode-locked oscillators [24,25] and more recently applied to other types of systems such as storage-rings [26] or free-electron lasers [27]. In brief, this procedure allows the discrete IMF [7] to be transformed into a simple differential equation using two timescales: $T$ describes the dynamics on the scale of the round-trip time, while $\theta$ resolves the pulse-shape itself. With this approach, the effects of dispersion, nonlinearity and loss/gain are modeled continuously around the ring cavity. One can show that the evolution of the envelope of the circulating electric field $A(T,\theta)$, normalized with respect to the square root of the pump pulse peak power $\sqrt{P_0}$, can be described by the following partial differential equation:

$$\left[\partial_T + \delta\partial_\theta\right]A(T,\theta) = -\alpha A - i/2\,\mathrm{sgn}(\beta_2)\partial_\theta^2 A$$
$$+ \sum_{k=3}^{\infty}\frac{i^{k+1}}{k!\,t_0^{k-2}}\frac{\beta_k}{\beta_2}\partial_\theta^k A + igF(A) + \xi A_p(\theta) \quad (1)$$

where $\delta = \tau\,t_0\,|\beta_2|^{-1}L^{-1}$ takes account of the temporal walk-off $\tau$ between the pump and the passive ring cavity, and the dimension-free fast timescale $\theta$ is defined in units of the characteristic pulse duration $t_0$. Second (higher) order dispersion is given by $\beta_2$ ($\beta_k$), $\alpha = \kappa T_R L_D/L$ is the amplitude loss/gain coefficient, $\kappa$ is the decay-rate of the photons inside the ring cavity, $T_R$ is the period of the pump-pulse train, $L_D = t_0^2/|\beta_2|$ is the dispersion length of the fiber and $L$ its physical length. The response $F(A)$, which can include for example Kerr, Raman, shock and nonlinear gain effects, is modelled by $gF(A)$, where $g = \gamma P_0 L_D$, with the usual nonlinear coefficient $\gamma$. The influence of the coherent pump is taken into account through the last term where $\xi = \sqrt{m}\left(\sqrt{R}L_D/L\right)$, $R$ represents the intensity reflectivity of the beam splitter, which couples the pump pulses into the cavity and $m = P/P_0$ is related to the peak power of the pump pulses that are injected into the cavity. $A_p(\theta)$ describes the arbitrary pump pulse envelope. In contrast to formally similar mean-field models [16,17,21,22,28,29], Eq. (1) allows for pump pulses with arbitrary temporal shapes and durations, as well as non-interferometric walk-off values.

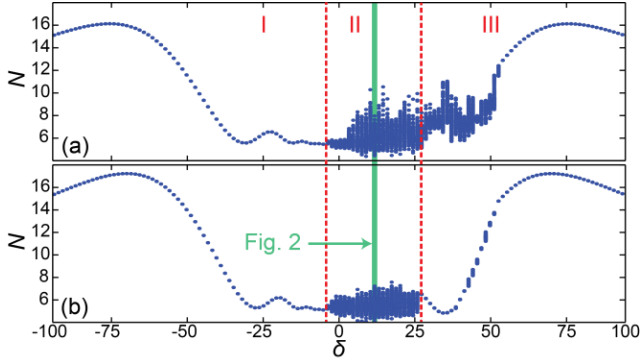

FIG. 1 (color online). Bifurcation diagrams of the norm $N$ with respect to the normalized delay $\delta$ corresponding to a delay range from −200 to 200 fs, simulated with (a) the IMF and (b) Eq. (1). Each value of $\delta$ contains 200 values of $N$. Vertical dashed lines represent dynamical transitions. The evolution diagrams in Fig. 2 correspond to the slice highlighted in green, which is labeled "Fig. 2".

We tested this model in a solid-core PCF ring cavity with characteristics similar to those used in recent experiments, including the full wavelength dependent dispersion [15]. The ring includes 20 cm of endlessly single mode PCF [30] pumped close to its zero dispersion wavelength, in the anomalous dispersion region. For simplicity, the analysis is limited to the instantaneous Kerr nonlinearity, $F(A) = |A|^2 A$, although Eq. (1) can be easily extended to include the Raman effect or any other kind of non-instantaneous nonlinear response. Fig. 1 shows the asymptotic behavior of this fiber ring cavity pumped by fs-pulses using both the IMF approach (Fig. 1a) and the new multiscale approach (Fig. 1b). For each value of $\delta$, the response of the cavity is modeled over 1200 loops for a train of Gaussian pump pulses ($t_0 = 100$ fs, peak power 1.5 kW). After discarding the transient response of the first 1000 loops, we plotted the norm $N = \int |A|^2 d\theta$, which is proportional to the physical pulse energy $E = P_0 t_0 N$ at every round trip. Three distinct regions can be identified: I and III show steady-state behavior ($N$ remains constant over all the plotted round-trips) while ii exhibits much more complex dynamics. The discrepancy in the position of the transition II → III can be explained by the fact that the transient regime becomes extremely long in the vicinity of $\delta = 28$. In this region the IMF has not yet fully converged to the final state after 1000 loops (Fig. 1a), while the multiscale model already has (Fig. 1b). In contrast, both models have already converged to the same final state at $\delta \approx -5$, where the transition I → II appears.

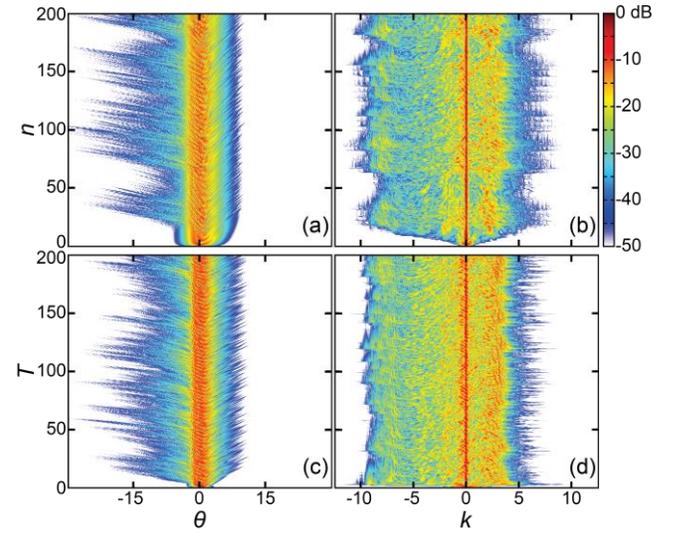

FIG. 2 (color online). Comparison of temporal (left column) and spectral (right column) pulse evolution simulated with the IMF (top) and Eq. (1) (bottom) for $\delta = 12$. The regime lies within the green region labeled "Fig. 2" in Fig. 1. The time window for $\theta$ corresponds to 6 ps in physical units and $k = 1/\theta$.

As a second test, the evolution of the temporal and spectral pulse shape inside the cavity is shown for $\delta \approx 12$, i.e., well inside the highly turbulent regime ii of Fig. 1. This value of $\delta$ corresponds to a walk-off value of 25 fs, i.e., much longer than an optical cycle ($\approx 3.5$ fs at a vacuum wavelength of 1042 nm). Both Fig. 1 and 2 show very good agreement between the commonly used discrete formalism and the multiscale model. It is also worth noticing that on the same processor, integration of the multiscale model is 5 times faster than the IMF.

We now use a reduced version of Eq. (1) in which only second order dispersion is considered. Experimentally, this could be realized using a kagomé-lattice hollow-core PCF filled with a high-pressure monatomic gas such as Xe [31]. Since monoatomic gases do not have any Raman response, Eq. (1) reduces to:

$$\partial_T A(\theta,T) = -\alpha A - i/2\,\mathrm{sgn}(\beta_2)\partial_\theta^2 A + i g\gamma |A|^2 A + \xi A_p(\theta)\ .\quad (2)$$

Note that Eq. (2) is *time-inversion symmetric* with respect to $\theta$. While the conventional approach to studying asymptotic dynamics in passive ring cavities is based on numerical pulse propagation, the multiscale method allows one to directly track the stationary states of Eq. (2) by imposing $\partial_T A = 0$. It can be straightforwardly solved using the numerical tool Auto [32]. We cross-checked the numerical solutions with an implementation of the Newton relaxation algorithm [33], as well as direct numerical pulse propagation.

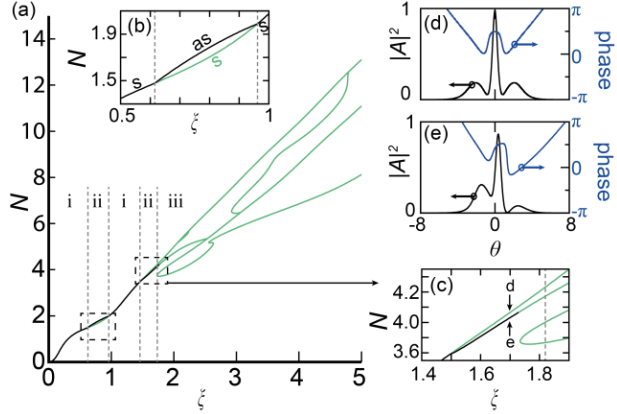

FIG. 3 (color online). (a) Bifurcation diagram of the norm $N$ of the stationary solutions of Eq. (2) with respect to the pulse strength $\xi$, where black branches are linearly stable and green branches unstable. i, ii, and iii indicate regions where the system is monostable, bistable and multistable. (b,c) Close-ups of the bistable and multistable regions. In (b), the black branch is completely symmetric (labeled "s") whereas the green branch is asymmetric (labeled "as"). (d,e) Temporal intensity and phase of the two solutions in the inset of (c). The vertical dashed line in (c) indicates the states discussed in Fig. 4.

Fig. 3a shows the complete bifurcation diagram of the norm $N$ of the stationary pulses circulating inside the cavity, plotted against the effective pump strength $\xi$ injected into the cavity. The pump wavelength lies in the anomalous dispersion regime. Note that in the normal dispersion regime the system does not bifurcate, but reduces to one unique branch similar to the upper branch in the case of anomalous dispersion. A standard linear stability analysis [33,34] was used to test the stability of the stationary states; stable states are drawn with black curves in Fig. 3, and unstable states with green curves. For the parameters of the ring cavity, the highest displayed injection parameter of $\xi = 5$ corresponds to a physical peak power of 5.8 kW.

Globally, Fig. 3a presents three distinct regions exhibiting different dynamics. As the effective pump strength is increased, the system bifurcates from monostable regions (i) to regions displaying bistability (ii) and even multistability (iii). At the same time the symmetry of the states changes in a complex way (Fig. 3). The spontaneous emergence of ASs is remarkable since Eq. (2) is time-inversion symmetric. Previous reports of ASs in fiber ring-cavities required the presence of odd-order dispersion [20], the Raman-effect or interferometric walk-off values [12]. Even more remarkable is that in the regions labeled ii of Fig. 3 the asymmetric state is stable against perturbations, whereas the symmetric solution is unstable. In region iii ($\xi > 1.74$) all the states are unstable, which coincides with the onset of multistability (Fig. 3c). In fact up to 5 solutions may coexist, both symmetric and asymmetric states being possible (the green branches in Fig. 3a). To the best of our knowledge, such multistable operation in a fiber ring cavity has never been reported before. In this regime, when the system is seeded with a randomly and weakly perturbed version of one of the unstable solutions of Eq. (2), it gradually evolves to a superposition of the possible eigenstates of the system.

Fig. 4 shows the simulated evolution of the circulating pulse when the system is seeded in this way with one of the symmetric states (the lowest one branch at $\xi = 1.83$ – see dashed line in Fig. 3d). The seed abruptly transits to the closest state (after $T \approx 145$), which then converges at $T \approx 450$ to a periodic beating between the steady-state solutions of Eq. (2) for this value of $\xi$. The initial change from the symmetric seed to the asymmetric steady-state solution of Eq. (2) can clearly be identified via the asymmetry parameter

$$\sigma = \int_{-\infty}^{0} |A(T,\theta) - A(T,-\theta)|^2\,d\theta / N\quad\text{(Fig. 4c). Note that for } T > 550$$

the asymmetry $\sigma$ oscillates between two values that do not correspond to isolated solutions of Eq. (2) (Fig. 4c).

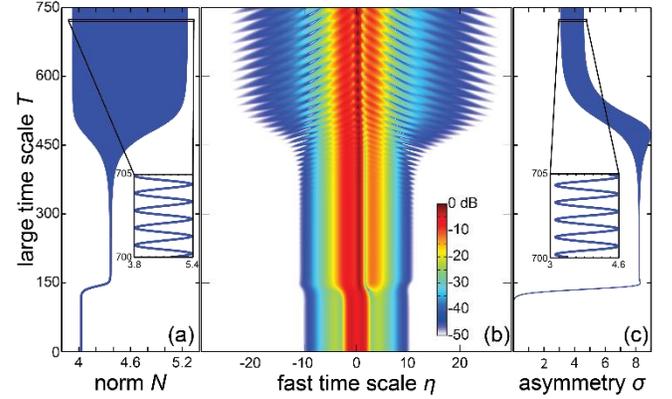

FIG. 4 (color online). Evolution of the (a) norm, (b) temporal shape and (c) asymmetry of the cavity pulse when seeding with an (unstable) symmetric solution at $\xi = 1.83$ as indicated by the vertical dashed line on Fig. 3d.

Furthermore, Eq. (2) does not preserve U(1) gauge symmetry [35], i.e., its solutions are not invariant under global phase transformations $A \to A\,e^{i\varphi}$, in contrast with other dissipative systems such as those described by Ginzburg-Landau (GL) equations [36,37]. In particular, U(1) symmetry is readily broken in Eq. (2) as a result of the interferometric superposition of the pump and the intracavity pulses. Among other implications, this remarkable feature of Eq. (2) compels any complex steady-state to be sensitive to global phase transformations and hence to display a unique phase portrait (Fig. 3(d,e)). In particular, the phase must be

"engineered" and preserved by the ring-cavity if the pulse shape is to be sustained.

In order to explain how the asymmetric states can be sustained within this temporally inversion-symmetric system, we employ a hydrodynamical analogy, based on the so-called Madelung transformation [38]. In brief, the optical field $A(T,\theta) = \sqrt{\rho}e^{i\Phi}$ is interpreted as a fluid, with density $\rho = |A|^2$ and velocity $v = \nabla\Phi(\theta)$ [39]. Similarly, we also introduce the general ansatz $A_p(\theta) = f(\theta)\,e^{i\psi}$ for the coherent pump in Eq. (2). Solving for the real and imaginary parts yields an Euler-like equation (real part) that accounts for conservation of momentum, and the following continuity-equation (imaginary part):

$$\partial_T \rho + \nabla J = 2\xi f(\theta)\sqrt{\rho}\cos(\psi - \phi) = S \qquad (3)$$

where $J = i/2\,(A\nabla A^* - A^*\nabla A) = \rho v$ is the energy flow and $S$ is the energy source term. As a result the pulse energy always flows from sources (regions with $S > 0$) to sinks (regions with $S < 0$). This means that for $J < 0$ ($J > 0$) energy flows towards the leading (trailing) edge. In general both $\Phi$ and $\psi$ depend on $\theta$ in Eq. (3). The coherent superposition of pump and cavity pulses manifests itself as a sinusoidal modulation of $S$, a feature that does not appear in systems (such as those governed by the GL equations) where power is supplied by internal amplification rather than a repetitive pump pulse train.

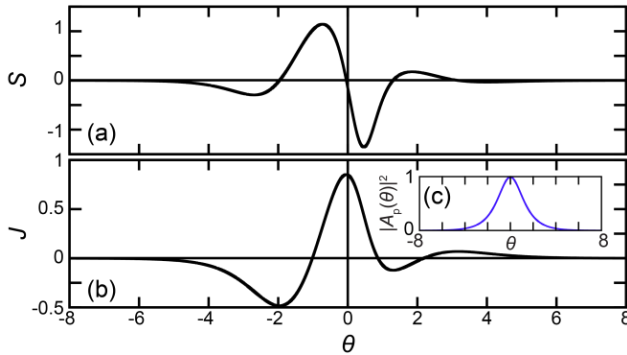

FIG. 5 (color online). Hydrodynamical representation of the (a) energy generation $S$ and (b) internal energy flow $J$ for the asymmetric state presented in Fig. 3e at $\xi \approx 1.7$. (c) shows the pump intensity $|A_p(\theta)|^2$.

It is worth noticing that any stationary state of the system, either temporally symmetric or asymmetric, fulfills energy conservation, i.e. $\int S dt = 0$.

Fig. 5 plots the energy generation $S$ and the energy flow $J$ for an asymmetric state corresponding to $\xi = 1.7$ (Fig. 3e). The redistribution of energy throughout the pulse structure, so as to compensate for losses and dispersion, is clear. Counterintuitively, the effective energy generation $S$ is almost negligible at $\theta = 0$ (Fig. 5a), even though the pump field (Fig. 5c) peaks at that point, as a result of constructive interference between pump and cavity pulses. Moreover, the positive energy flow leads to a redistribution of the energy towards positive $\theta$ (i.e. the pulse tail). This explains why the peak intensity of this stationary pulse does not overlap with the center of the pump pulse, but is shifted towards $\theta > 0$ (Fig. 3c).

In conclusion, we have introduced a new multiscale model for studying the dynamics of fiber-ring cavities. The model can handle any order of dispersion, as well as temporal walk-off values much longer than an optical cycle. Moreover, the model is not limited to continuous pumping but can be used for arbitrary pump pulse shapes and durations. In this Letter we have explored the dynamics in the ultra-short pumping regime (~100 fs). It is, however, straightforward to check whether similar bifurcations from symmetric to asymmetric states can appear for longer pulses (~10 ps). We have successfully tested the model against the well-established and exact Ikeda-map formalism, with excellent agreement. The equation is also computationally much faster than the Ikeda approach. A Raman-free version of the model shows (surprisingly) that fiber-ring cavities described by time-inversion symmetric models can exhibit spontaneous symmetry breaking as well as multistability. Such a system, with very weak dispersion, is experimentally feasible if kagomé-PCF filled with high-pressure monoatomic gas is used as the nonlinear element. This new multiscale model, and the hydrodynamics approach used to explain how the asymmetric states can be self-sustained, should prove useful for modeling the behavior of arbitrary nonlinear waveguide-based ring cavities. The generality of the approach and the governing equation may be useful to a broad range of communities across many different fields.

We are grateful to Serge Bielawski and Christophe Szwaj for fruitful discussions.

[1] L. D. Landau and E. M. Lifshitz, *Statistical Physics* (Elsevier, 1996).
[2] P. W. Higgs, Phys. Rev. Lett. **13**, 508 (1964).
[3] A. E. Miroshnichenko, B. A. Malomed, and Y. S. Kivshar, Phys. Rev. A **84**, 012123 (2011).
[4] Y. Li, J. Liu, W. Pang, and B. A. Malomed, Phys. Rev. A **87**, 013604 (2013).
[5] J. R. Salgueiro and Y. S. Kivshar, Opt. Lett. **30**, 1858 (2005).
[6] C. Green, G. B. Mindlin, E. J. D'Angelo, H. G. Solari, and J. R. Tredicce, Phys. Rev. Lett. **65**, 3124 (1990).
[7] I. Kensuke, Opt. Commun. **30**, 257 (1979).
[8] H. U. Voss, A. Schwache, J. Kurths, and F. Mitschke, Phys. Lett. A **256**, 47 (1999).
[9] R. Vallée, Opt. Commun. **81**, 419 (1991).
[10] G. Steinmeyer and F. Mitschke, Appl. Phys. B Laser Opt. **62**, 367 (1996).
[11] J. García-Mateos, F. C. Bienzobas, and M. Haelterman, Fiber Integr. Opt. **14**, 337 (1995).
[12] F. Leo, L. Gelens, P. Emplit, M. Haelterman, and S. Coen, Opt. Express **21**, 9180 (2013).
[13] P. St.J. Russell, J. Light. Technol. **24**, 4729 (2006).

[14] N. Brauckmann, M. Kues, P. Groß, and C. Fallnich, Opt. Express **19**, 14763 (2011).
[15] M. Schmidberger, W. Chang, P. St.J. Russell, and N. Y. Joly, Opt. Lett. **37**, 3576 (2012).
[16] L. A. Lugiato and R. Lefever, Phys. Rev. Lett. **58**, 2209 (1987).
[17] F. Leo, S. Coen, P. Kockaert, S.-P. Gorza, P. Emplit, and M. Haelterman, Nat. Photonics **4**, 471 (2010).
[18] M. Kues, N. Brauckmann, P. Groß, and C. Fallnich, Phys. Rev. A **84**, 033833 (2011).
[19] M. J. Schmidberger, F. Biancalana, P. St.J. Russell, and N. Y. Joly, in *Eur. Conf. Lasers Electro-Opt.* (OSA, Munich, Germany, 2013).
[20] F. Leo, A. Mussot, P. Kockaert, P. Emplit, M. Haelterman, and M. Taki, Phys. Rev. Lett. **110**, 104103 (2013).
[21] G. Kozyreff, M. Tlidi, A. Mussot, E. Louvergneaux, M. Taki, and A. G. Vladimirov, Phys. Rev. Lett. **102**, 043905 (2009).
[22] M. Tlidi, A. Mussot, E. Louvergneaux, G. Kozyreff, A. G. Vladimirov, and M. Taki, Opt. Lett. **32**, 662 (2007).
[23] Y. K. Chembo and C. R. Menyuk, Phys. Rev. A (2013).
[24] H. A. Haus, IEEE J. Quant. Electron. **6**, 1173 (2000).
[25] T. Kolokolnikov, M. Nizette, T. Erneux, N. Y. Joly, and S. Bielawski, Physica D **219**, 13 (2006).
[26] P. Elleaume, IEEE J. Quant. Electron. **21**, 1012 (1985).
[27] C. Bruni, T. Legrand, C. Szwaj, S. Bielawski, and M. E. Couprie, Phys. Rev. A **84**, 063804 (2011).
[28] S. Coen, H. G. Randle, T. Sylvestre, and M. Erkintalo, Opt. Lett. **38**, 37 (2013).
[29] M. Tlidi and L. Gelens, Opt. Lett. **35**, 306 (2010).
[30] T. A. Birks, J. C. Knight, and P. S. J. Russell, Opt. Lett. **22**, 961 (1997).
[31] M. Azhar, N. Y. Joly, J. C. Travers, and P. St.J. Russell, Appl. Phys. B 1 (2013).
[32] E. Doedel, H. B. Keller, and J. P. Kernevez, Int. J. Bifurc. Chaos **01**, 745 (1991).
[33] A. Alexandrescu and J. R. Salgueiro, Comput. Phys. Commun. **182**, 2479 (2011).
[34] J. M. Soto-Crespo, D. R. Heatley, E. M. Wright, and N. N. Akhmediev, Phys. Rev. A **44**, 636 (1991).
[35] C. Sulem and P.-L. Sulem, *The Nonlinear Schrödinger Equation: Self-Focusing and Wave Collapse* (Springer, 1999).
[36] N. Akhmediev and A. Ankiewicz, in *Dissipative Solitons*, edited by N. Akhmediev and A. Ankiewicz (Springer Berlin Heidelberg, 2005), pp. 1–17.
[37] I. S. Aranson and L. Kramer, Rev. Mod. Phys. **74**, 99 (2002).
[38] E. Madelung, *Die mathematischen Hilfsmittel des Physikers* (Springer, 1964).
[39] D. Novoa, H. Michinel, and D. Tommasini, Phys. Rev. Lett. **103**, 023903 (2009).